\documentclass[epj]{webofc}
\usepackage[varg]{txfonts}   % Web of Conferences font
\usepackage{graphicx,color} % for figures
\usepackage{bm}       % bold math 
\usepackage{amsmath}  
\usepackage{amssymb}
\woctitle{21st International Conference on Few-Body Problems in Physics}
\begin{document}
\title{Towards a 
Theoretical Understanding
of the $\bm{XYZ}$  Mesons \\
from QCD}
\author{Eric Braaten\inst{1}\fnsep\thanks{\email{braaten@mps.ohio-state.edu}} 
%\and
%        Second author\inst{2}\fnsep\thanks{\email{secondauthor@alab.gov}} \and
   %     Third author\inst{3}\fnsep\thanks{\email{Emailaddressforthirdauthorifnecessary}}
        % etc.
}

\institute{Department of Physics, Ohio State University,
Columbus OH 43210 USA}
%\and
%           the second here 
%\and
%           Last address
 %         }

\abstract{
%  Insert your abstract here. 
The $XYZ$ mesons are mesons that contain a heavy quark and antiquark 
but have properties that seem to require additional constituents.
Some of them are electrically charged,
so they must be tetraquark mesons
whose additional constituents are a light quark and antiquark.
The list of $XYZ$ mesons has grown to about two dozen over the last decade.
A promising approach to understanding these mesons within QCD
is the Born-Oppenheimer approximation,
which reduces the problem to the solution of the Schr\"odinger equation 
in potentials that can be calculated using lattice QCD.
The Born-Oppenheimer approximation has not yet revealed the pattern of the $XYZ$ mesons,
but it provides a compelling framework for understanding them from the fundamental theory.
}
\maketitle
\section{Introduction}
\label{intro}
%Your introduction comes here. Then, separate text sections with

The $XYZ$  mesons are more than two dozen new $c \bar c$ and $b \bar b$ mesons
that have been discovered since 2003  \cite{Bodwin:2013nua}.
They contain a heavy quark and antiquark,
but they have properties that seem to require additional constituents.
Some of the more remarkable $XYZ$ mesons are
\begin{itemize}
\item
$X(3872)$, 
discovered by the Belle Collaboration in 2003 \cite{Choi:2003ue}.  
It has comparable branching fractions into $J/\psi\, \rho$ 
and $J/\psi\, \omega$, implying a severe violation of isospin symmetry.
\item
$Y(4260)$, 
discovered by the BaBar Collaboration in 2005 \cite{Aubert:2005rm}.  
It has $J^{PC}$ quantum numbers $1^{--}$, but it is produced very weakly 
in $e^+e^-$ annihilation.
\item
$Z^+(4430)$, 
discovered by the Belle Collaboration in  2007 \cite{Choi:2007wga}.  
It decays into $\psi(2S)\, \pi^+$, which implies that it must be a tetraquark meson 
with constituents $c\bar c u \bar d$.
\item
$Y(4140)$, 
discovered by the CDF Collaboration in 2009 \cite{Aaltonen:2009tz}.  
It decays into $J/\psi\, \phi$, which suggests that it could be a tetraquark 
meson with constituents $c \bar c s \bar s$.
\item
$Z_b^+(10610)$ and $Z_b^+(10650)$, 
discovered by the Belle Collaboration in 2011 \cite{Belle:2011aa}.  
They both decay into $\Upsilon\, \pi^+$, 
which implies that they must be tetraquark mesons with constituents 
$b\bar b u \bar d$.
\item
$Z_c^+(3900)$, 
discovered by the BESIII Collaboration in 2013 \cite{Ablikim:2013mio}.  
It decays into $J/\psi\, \pi^+$, which implies that it must be a tetraquark meson 
with constituents $c\bar c u \bar d$.
\end{itemize}
Many of  the $XYZ$  mesons are surprisingly narrow.
They present  a major challenge to our understanding of the QCD spectrum!

\section{Models for $\bm{XYZ}$ Mesons}
\label{sec-1}
%For bibliography use \cite{RefJ}

Most of the theoretical work on the  $XYZ$ mesons
has been carried out using constituent models.
These models fall into three basic categories:
(1) {\it conventional quarkonium},
(2) {\it quarkonium hybrids},
(3) {\it quarkonium tetraquarks}.

In a conventional quarkonium, 
the only constituents are a heavy quark $Q$ and antiquark $\bar Q$.
There is a well-developed phenomenology for  conventional quarkonium
based on quark potential models.
These models are accurate below the open-heavy-flavor threshold,
and they give well-defined predictions above the threshold.
They imply that the heavy quarkonium states have energy levels 
labeled by radial and orbital-angular-momentum quantum numbers $nL$.
Each energy level consists of  a heavy-quark-spin multiplet
that includes a spin-singlet state and one or three spin-triplet states
with specific $J^{PC}$ quantum numbers.
For example, an S-wave multiplet consists of the two states $\{0^{-+}, 1^{--} \}$,
a  P-wave multiplet consists of the four states $\{1^{+-}, (0,1,2)^{++}\}$,
and a  D-wave multiplet consists of the four states $\{2^{-+}, (1,2,3)^{--}\}$.
There are some exotic quantum numbers that are not allowed for quarkonium. 
If the only constituents are a quark and an antiquark,
the forbidden quantum numbers are $0^{--}$, $0^{+-}$, $1^{-+}$, $2^{+-}$, $3^{-+}$, \ldots.

In a quarkonium hybrid, the $Q$ and $\bar Q$ are accompanied by a 
gluonic excitation.  There are various models for the gluonic excitation.
In a {\it constituent gluon model}, it is a massive spin-1 particle.
In a {\it flux tube model}, it is a vibrational state of a color flux tube
extending between the heavy quark and antiquark. 
In a {\it gluelump model}, it is gluon fields bound to a
compact $Q \bar Q$ pair in a color-octet state.

In a quarkonium tetraquark, the $Q$ and $\bar Q$ are accompanied by a 
light quark $q$ and a light antiquark $\bar q$.  
There are many ways the four colored constituents can be clustered in the meson.
In a {\it compact tetraquark}, they are all in overlapping orbitals. 
In a {\it meson molecule}, there are two color-singlet clusters  $Q \bar q$ and $\bar Q q$.
In a {\it diquarkonium}, the constituents are clustered into 
color-triplet diquarks $Q q$ and $\bar Q \bar q$.
In a {\it hadro-quarkonium}, the $Q \bar Q$ pair forms a compact color-singlet core
to which the $q$ and $\bar q$ are bound.
In a {\it quarkonium adjoint meson}, the $Q \bar Q$ pair forms a compact color-octet core
to which the $q$ and $\bar q$ are bound.

The various constituent models for $XYZ$ mesons make little connection with 
the fundamental theory QCD.
The constituents are plausible degrees of freedom from QCD,
but the interactions between them are purely phenomenological.
They have had some success in describing individual $XYZ$ mesons,
but they have failed to reveal the pattern of $XYZ$ mesons.
These models can appropriately be compared to
blind-folded physicists studying an elephant.
Depending on what aspect of the  elephant they focus on, 
whether it be an ear or a tusk or the trunk or the tail,
they arrive at very different conclusions about the nature of an elephant.
All of these conclusion have some element of truth,
but none of them comes close to capturing the true nature of an elephant.

In order to understand the true nature of the elephant,
it is essential to approach the problem from within QCD.
The starting point should be the fundamental fields, which are quarks and gluons.
It should be expressed in terms of the parameters of QCD,
which are the coupling constant $\alpha_s$ and the quark masses.
Three approaches that fulfill these requirements are {\it QCD sum rules},
{\it lattice QCD}, and the {\it Born-Oppenheimer approximation}.
I will not discuss {QCD sum rules, 
because they seem to be too blunt a method to be very useful for this problem.
I will summarize what is known from lattice QCD.
I will then turn to the Born-Oppenheimer approximation,
which I believe is the key to understanding the  $XYZ$ mesons.

\section{Lattice QCD}
\label{sec-2}

The most extensive calculations of the  $c \bar c$ meson spectrum using lattice gauge theory
were pioneered by Dudek, Edwards, Mathur, and Richards  \cite{Dudek:2007wv}
and extended by the Hadron Spectrum Collaboration \cite{Liu:2012ze}.
They used  lattice QCD on an anisotropic lattice with $24^3 \times128$ sites,
but with light quarks that were too heavy.
Their pion mass was 400~MeV, which is about 3 times its physical value.
The most important caveats on their results are that 
they made no extrapolation to the physical up and down quark masses   
and they made no extrapolation to zero lattice spacing.
Their results were impressive nonetheless.
They determined the  masses of charmonium and charmonium hybrids
from the cross-correlators of many operators.
They identified 46 statistically significant states with various $J^{PC}$
quantum numbers, with spin $J$ as high as 4 and masses as high as 4.6~GeV.
They were also able to discriminate between charmonium and charmonium hybrids 
based on how strongly the states coupled to operators 
that had a factor of the gluon field strength.
They identified 14 charmonium hybrid candidates. 
The 4 lowest candidates fill out a complete heavy-quark-spin multiplet:
$\{1^{--}, (0,1,2)^{-+} \}$.
One of these quantum numbers is exotic, namely $1^{-+}$.
Their other charmonium hybrid candidates were 3 spin singlets, $(0,1,2)^{++}$, 
and 7 spin triplets, $(0,1,1,1,2,2,3)^{+-}$.
Two of these quantum numbers are exotic, namely $0^{+-}$ and $2^{+-}$.

Lattice gauge theory has also been applied to the  $b \bar b$ meson spectrum.
Since the $b$ quark is three times as heavy as the $c$ quark,
it is more practical to use a nonrelativstic effective field theory called NRQCD.
The most extensive calculations of the  $b \bar b$ meson spectrum using lattice NRQCD
are still the pioneering calculations  by Juge, Kuti, and Morningstar back in 1999 \cite{Juge:1999ie}.
They used quenched  lattice NRQCD on an anisotropic lattice with $15^3 \times 45$ sites.
The adjective ``quenched'' implies that there were no light quarks at all,
which necessarily introduces large systematic errors.
They  identified 4 bottomonium hybrid states,
all spin singlets with quantum numbers $1^{--}$, $1^{++}$, $0^{++}$, 
and a second $1^{--}$.
The quantum numbers $1^{--}$ of the lowest state match those of the spin-singlet member of the 
lowest multiplet of charmonium hybrid candidates 
in the lattice QCD calculations of the Hadron Spectrum Collaboration.
The quantum numbers $1^{++}$ and $0^{++}$ of the next two states 
match those of two of the other three spin-singlet  charmonium hybrid candidates.

There have been some applications of lattice gauge theory 
to charmonium tetraquark mesons,
which have the flavor of a light quark and a light antiquark.
Prelovsek, Leskovec, and collaborators
used lattice QCD on an anisotropic lattice with $16^3 \times 32$ sites,
but with two light quarks that were too heavy \cite{Prelovsek:2013cra,Prelovsek:2013xba,Prelovsek:2014swa}.
Their pion mass was 270~MeV, which is about twice its physical value.
They studied only a few specific $J^{PC}$ channels,
and used a rather small set of  interpolating operators.
In the $J^{PC}= 1^{++}$ channel, they found a candidate for the $X(3872)$,
 but it is probably just the $2P$ charmonium state $\chi_{c1}(2P)$.
In the $J^{PC} = 1^{+-}$ channel, they found no signal for a $Z_c^+$ tetraquark.

There have not yet been any applications of lattice gauge theory 
to bottomonium tetraquark mesons.

\section{Born-Oppenheimer Approximation
         for Quarkonium Hybrids}
\label{sec-3}

The Born-Oppenheimer approximation in atomic physics 
provides a framework for understanding molecules
and the low-energy scattering of atoms \cite{Born-Oppenheimer}.
The Born-Oppenheimer approximation for QCD was
pioneered by Juge, Kuti, and Morningstar,
who applied it to quarkonium and quarkonium hybrids  \cite{Juge:1999ie}.
It is based on the fact that the
heavy-quark mass $m_Q$ is much larger than the
energy scale for gluons.
Because they are heavy, the $Q$ and $\bar Q$ in a 
 quarkonium ,or in a quarkonium hybrid, move nonrelativistically.
The gluons can respond almost instantaneously 
to the motion of the $Q$ and $\bar Q$.
Given the positions of the $Q$ and $\bar Q$, 
the gluon fields will be in a stationary state
in the presence of static $Q$ and $\bar Q$ sources at those positions.     
As the positions of the $Q$ and $\bar Q$ change, 
 the gluon fields will usually remain adiabatically in that stationary state.
The energy of a stationary state of gluon fields
 in the presence of static $Q$ and $\bar Q$ sources
separated by the distance $R$  
defines a Born-Oppenheimer potential $V(R)$.                          
In the simplest Born-Oppenheimer approximation,
the motion of the $Q \bar Q$ pair is described by 
the Schr\"odinger equation in the potential $V(R)$.                        

The stationary states for gluon fields
in the presence of static $Q$ and $\bar Q$ sources
 can be labeled by the quantum numbers 
 that are conserved in the presence of the sources.
 A convenient set of quantum numbers are
 \begin{itemize}
 \item
the absolute value $| \bm{\hat R} \cdot \bm{J}_{\text{light}}|$
of the component of the angular momentum of the light fields
along the axis defined by the sources.
It has integer values $0, 1, 2, \ldots$,
but they are traditionally labeled by upper case Greek letters $\Sigma, \Pi, \Delta, \ldots$,
\item
the product $(CP)_{\text{light}}$ of the charge conjugation and parity
of the light fields.  Its values are $+1$ or $-1$,
but they are traditionally labeled by a subscript $g$ or $u$,
\item
in the case of $\Sigma$, an additional quantum number for a reflection 
through a plane containing the sources.  Its values are $+1$ or $-1$,
but they are traditionally labeled by a superscript $+$ or $-$ on the $\Sigma$.
\end{itemize}
Thus the Born-Oppenheimer potentials are labeled 
$\Lambda_\eta$, where
$\Lambda =\Sigma^+, \Sigma^-, \Pi, \Delta, \ldots$
and $\eta = g,u$.  

\begin{figure}
\centering
\includegraphics[width=10cm,clip]{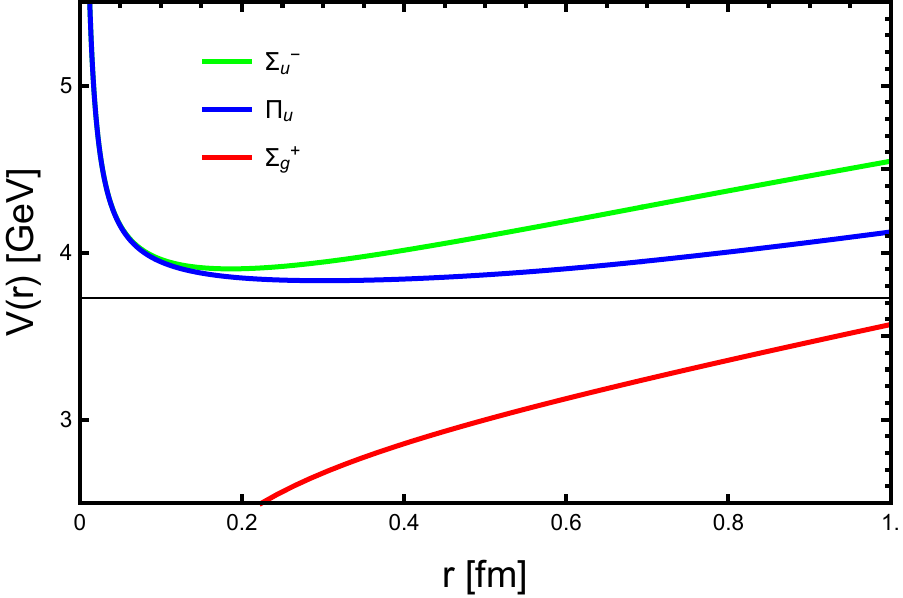}
\caption{The deepest flavor-singlet Born-Oppenheimer potentials:
the quarkonium potential $\Sigma_g^+$
and the quarkonium hybrid potentials $\Pi_u$ and $\Sigma_u^-$.
An additive constant in the potentials has been chosen 
so that the energy levels are those of $c \bar c$ mesons.
The thin horizontal line is the charm-meson-pair threshold.
For $b \bar b$ mesons, the potentials are the same 
except for an additional energy offset of  about 6.8 GeV.}
\label{fig-1}       % Give a unique label
\end{figure}

Many of the Born-Oppenheimer potentials were calculated 
by Juge, Kuti, and Morningstar in 1999 using quenched lattice QCD
on an anisotropic lattice with $10^3 \times 30$ sites   \cite{Juge:1999ie}.
The deepest Born-Oppenheimer potentials are
$\Sigma_g^+$, then $\Pi_u$, and then $\Sigma_u^-$.  
These potentials are illustrated in Figure~\ref{fig-1}.                       
Juge, Kuti, and Morningstar also
solved the Schr\"odinger equation in the Born-Oppenheimer potentials.
The energy levels are labeled by $nL$,
where $n = 1,2,3,\ldots$ is a radial quantum number
and $L= 0,1,2,\ldots$  (or $S,P,D,\ldots$)
is an orbital-angular-momentum quantum number.                            
The energy levels in the $\Sigma_g^+$ potential
are quarkonium states.
The energy levels in the excited potentials,
including $\Pi_u$ and $\Sigma_u^-$, are quarkonium hybrids.
Juge, Kuti, and Morningstar found that the 
predictions of the Born-Oppenheimer approximation for the energy splittings
between the bottomonium hybrid energy levels were in semiquantitative agreement 
with their direct calculations using quenched lattice NRQCD   \cite{Juge:1999ie}.
This convinced me back in 1999 that quarkonium hybrids are real states in the QCD spectrum.                          
                           
The lowest charmonium hybrid energy levels predicted by the 
Born-Oppenheimer approximation are 
\begin{itemize}
\item
the $\Pi_u(1P)$ energy level,
which consists of the two heavy-quark-spin multiplets
$\{1^{--},  (0,1,2)^{-+}\}$ and $\{1^{++},  (0,1,2)^{+-}\}$,
\item
the $\Pi_u(1D)$ energy level,
which consists of the two multiplets
$\{2^{--},  (1,2,3)^{-+}\}$ and $\{2^{++},  (1,2,3)^{+-}\}$,
\item
the $\Sigma_u^-(1S)$ energy level,
which consists of the multiplet $\{0^{++},  1^{+-}\} $.                         
\end{itemize}
 The lowest multiplet $\{1^{--},  (0,1,2)^{-+}\}$    
matches the lowest multiplet of charmonium hybrid candidates
in the lattice QCD calculations of the Hadron Spectrum Collaboration  \cite{Liu:2012ze}.
Their other 10 charmonium hybrid candidates have quantum numbers that match
all the states in three of the next 4 highest Born-Oppenheimer energy levels.
Thus lattice QCD seems to reproduce the pattern of states predicted by the Born-Oppenheimer approximation,
even when the light quarks are too heavy.
 
We proceed to discuss the  
Born-Oppenheimer potentials at small $R$. 
 The lowest hybrid potentials are $\Pi_u$ and then $\Sigma_u^-$.
At small $R$, these potentials increase roughly like $1/R$.
The $Q$ and $\bar Q$ sources behave like a local color-octet source
and therefore like a gluino, which is a hypothetical heavy particle  predicted by supersymmetry
that has the same color-octet charge as the gluon.
The stationary state of gluon fields bound to a color-octet source
(or a gluino) is called a {\it gluelump}.
 The potential at small $R$ is the  repulsive Coulomb potential
 between the $Q$ and $\bar Q$ plus the energy of the gluelump.
The gluelump spectrum has been calculated using lattice QCD.
The most recent calculations were by Marsh and Lewis 
using an anisotropic lattice with $28^3 \times 56$ sites,
but with light quarks that were too heavy \cite{Marsh:2013xsa}.
Their pion mass is $m_\pi = 480$~MeV,
which is about 3 times its physical value.
The lowest-energy gluelump has  quantum numbers $1^{+-}$.
The next gluelump has quantum numbers $1^{--}$ 
and energy larger by about 300~MeV.     
The energy of the $1^{+-}$ gluelump  determines the common  limit as $R \to 0$
of the  $\Pi_u$ and $\Sigma_u^-$ potentials.               
Thus, just by knowing the quantum numbers $1^{+-}$ of the lowest-energy gluelump,
one can infer correctly that the deepest hybrid Born-Oppenheimer potentials                           
are probably $\Pi_u$ and $\Sigma_u^-$.
                       
We now discuss the  
Born-Oppenheimer potentials at large $R$.                      
If there are no light quarks,
all the  Born-Oppenheimer potentials  increase linearly with $R$ at large $R$.
The corresponding stationary state is a color flux tube 
extending between the $Q$ and $\bar Q$ sources.                        
The asymptotic behavior of the potential at large $R$ is $V(R) \to \sigma R$, 
where $\sigma$ is the energy per length of the color flux tube.                       
If there are light quarks,
the lowest-energy stationary state  at large $R$ actually consists of 2 static mesons,
one consisting of a light antiquark bound to the  $Q$ source                      
and the other consisting of a light quark bound to the  $\bar Q$ source.                     
In this case, the potential approaches a constant at large $R$
that is equal to twice the energy of a static meson.
One might expect that the linearly-rising quarkonium and hybrid potentials   
cease to exist above the energy of the static-meson-pair threshold.                    
However stationary states with  linearly-rising potentials 
do exist at higher energies, but they have avoided crossings with the threshold.
This has been shown convincingly for the quarkonium $\Sigma_g^+$ potential
by the SESAM Collaboration using lattice QCD  \cite{Bali:2005fu}.
They used an anisotropic lattice with $24^3 \times 40$ sites,
but with light quarks that are too heavy.
Their pion mass was 540~MeV, which is about 4 times its physical value.                                              
This work demonstrated that avoided crossings between 
Born-Oppenheimer potentials can be calculated using lattice QCD.

 \section{Born-Oppenheimer Approximation
         for Quarkonium Tetraquarks}
\label{sec-3}

Those $XYZ$ mesons that have an electric charge are definitely quarkonium tetraquarks
whose constituents include a $Q \bar Q$ pair  and a light quark and antiquark.
Some of the neutral $XYZ$ mesons may also be quarkonium tetraquarks.
Quarkonium tetraquarks can be treated using the Born-Oppenheimer approximation 
just like quarkonium hybrids  \cite{Braaten:2013boa}.
This is because light quarks will respond almost instantaneously to the motion of the heavy quarks, 
just like gluon fields.                
The difference is that the stationary state of gluons and light quarks 
in a quarkonium tetraquark
not only has Born-Oppenheimer quantum numbers, but also light-quark+antiquark flavors.
Unfortunately there are as yet  no lattice QCD calculations 
of tetraquark Born-Oppenheimer potentials.

There is one hint  from lattice QCD about the tetraquark Born-Oppenheimer potentials, 
and that comes from calculations of the {\it adjoint meson} spectrum. 
An adjoint meson consists of light-quark and gluon fields 
with light-quark+antiquark flavor that are bound to a color-octet source (or a gluino).                
The adjoint meson spectrum was calculated by Foster and Michael                 
using quenched lattice QCD 
on an anisotropic lattice with $24^3 \times 48$ sites \cite{Foster:1998wu}.
They found that the quantum numbers of the  lowest-energy adjoint meson 
was  $1^{--}$  or  $0^{-+}$.
The energy of the $1^{--}$ adjoint meson determines the common  limit as $R \to 0$
of the  $\Pi_g$ and $\Sigma_g^+$ potentials.               
The energy of the $0^{-+}$ adjoint meson determines the limit as $R \to 0$
of the  $\Sigma_u^-$ potential.               
Thus, if the quenched lattice QCD calculations have correctly identified 
the quantum numbers of the lowest-energy adjoint mesons,
one can infer that the deepest tetraquark Born-Oppenheimer potentials
are probably $\Pi_g$ and $\Sigma_g^+$ or $\Sigma_u^-$.
The Born-Oppenheimer potentials may be different for each of three light-quark+antiquark flavors:  
isospin 1, isospin 0 and $s \bar s$.

The energy levels in the tetraquark Born-Oppenheimer potentials
are labeled by quantum numbers $nL$.
If the quenched lattice QCD calculations have correctly identified the 
the quantum numbers of the lowest-energy adjoint mesons,
some of the lowest energy levels are likely to be
\begin{itemize}
\item
the $\Pi_u(1P)$ energy level,
which consists of the heavy-quark-spin multiplets
$ \{1^{-+},  (0,1,2)^{--}\}$ and $ \{1^{+-},  (0,1,2)^{++}\}$,
\item
the $\Sigma_g^+(1S)$ energy level,
which consists of the multiplet $\{0^{-+},  1^{--}\}$,
\item
the $\Sigma_u^-(1S)$ energy level,
which consists of the multiplet $\{0^{++},  1^{+-}\}$.
\end{itemize}
There will be Born-Oppenheimer energy levels such as these for each of three light-quark+antiquark flavors:  
isospin 1, isospin 0 and $s \bar s$.
Thus there are many possibilities for the $J^{PC}$ and flavor
quantum numbers of the quarkonium tetraquarks.

 \section{$\bm{XYZ}$ Mesons from the Born-Oppenheimer Perspective}
\label{sec-4}

The various constituent models for the $XYZ$ mesons are based on 
different assumptions about their nature.
What is their nature from the Born-Oppenheimer perspective?
Are they compact tetraquarks?
Are they diquarkonium?
Are they adjoint mesons?
Are they meson molecules?
The answer is all of the above!
Each of these possibilities describes some region    
of the Born-Oppenheimer wavefunction.
In the small-$R$ region where the hybrid potentials are repulsive Coulomb potentials,
the $Q \bar Q$ pair form a compact color-octet source,
so the configuration is like a gluelump or an adjoint meson.
In the large-$R$ region where the potential increases linearly,
there is a color flux tube connecting a diquark
consisting of the light quark bound to the $Q$ 
and another diquark consisting of
the light antiquark bound to the $\bar Q$,
so the configuration is like a diquarkonium.
In the intermediate region,
so the configuration is like a compact tetraquark.
In the region of $R$ beyond an avoided crossing,
the configuration is two separated mesons, so it is like a meson molecule.

The  Born-Oppenheimer approach implies that
quarkonium hybrids and quarkonium tetraquarks definitely exist as states in the QCD spectrum.      
Whether they can be observed in experiments depends on 
how narrow they are, how easily they can be produced,
and whether they have favorable decay modes.        
The Born-Oppenheimer approximation
has not yet revealed a compelling pattern for the $XYZ$ mesons
There are too many unknown Born-Oppenheimer potentials
and too few $XYZ$ mesons with known $J^{PC}$ quantum numbers.
Selection rules for hadronic transitions
between Born-Oppenheimer configurations have been derived
 \cite{Braaten:2014ita,Braaten:2014qka}.
They provide useful constraints, but these constraints have proven 
to be insufficient to reveal the pattern of the $XYZ$ mesons.

 \section{Conclusions}
\label{sec-4}

The discoveries of the $XYZ$ mesons
have revealed a serious gap in our understanding of the QCD spectrum.
Constituent models for the $XYZ$ mesons
have not presented a compelling pattern, and they make little contact with QCD.
The Born-Oppenheimer approximation 
has not yet provided a compelling pattern for the $XYZ$ mesons,
but it is based firmly on QCD.                  

To fully develop the  Born-Oppenheimer approximation                  
would require  a lot of information from lattice QCD.
One simple problem is the calculation of the spectrum of gluelumps and adjoint mesons, 
which determines the small-$R$ behavior of the Born-Oppenheimer potentials.
It would be useful to have calculations of the $b\bar b$ hybrid meson spectrum 
with quality comparable to that of the  $c\bar c$ hybrid meson spectrum
calculated by  the Hadron Spectrum Collaboration. 
Some more challenging problems are calculations of the avoided crossings between 
the Born-Oppenheimer potentials and meson-pair thresholds
and calculations of the tetraquark Born-Oppenheimer potentials.

There are also many things that are needed from non-lattice theory.
In the absence of lattice QCD calculations of the Born-Oppenheimer potentials,
it might be possible to infer them from data on $XYZ$ mesons.
After all, the quarkonium potential was inferred quite accurately 
from very limited data on the charmonium spectrum way back in 1975 \cite{Eichten:1974af}.
It would be useful to develop quantitative phenomenological models for hadronic transitions 
between quarkonium, quarkonium hybrids,  and quarkonium tetraquarks,
because most of the $XYZ$ mesons have been observed through hadronic transitions.
Finally, it would be desirable to develop an effective field theory
in which the Born-Oppenheimer approximation arises as a first approximation,
so that corrections can be calculated  systematically.

Theory has so far failed to reveal the pattern of the $XYZ$ mesons.
Fortunately, future experimental progress on the $XYZ$ mesons  is guaranteed.
There are ongoing $e^+e^-$ experiments at the charm factory in China,
and upcoming $e^+e^-$ experiments at the bottom factory Super KEK-B in Japan.
In the ongoing $p p$ collision experiments at the LHC,
the ATLAS and CMS detectors can contribute to the study of $XYZ$ mesons
and the LHCb detector is especially well-suited for this purpose.
Finally, on the horizon, there is the PANDA experiment,  which will use $p \bar p$ annihilation at resonance
to produce charmonium, charmonium hybrids, and charmonium tetraquarks.
What is needed from experiment is more $J^{PC}$ quantum numbers,
more transitions (hadronic and radiative), and more $XYZ$ mesons.
With enough new information and with the Born-Oppenheimer approximation as a guiding principle,
the solution to the puzzle of the $XYZ$ mesons is inevitable.             
 
\begin{acknowledgement}
This work was supported in part by the Department of Energy,
by the National Science Foundation, and by the Simons Foundation.
%The acknowledgement environment is usually used as the last paragraph to acknowledge
%financial support from agencies and say nice things about collaborators---if appropriate.
\end{acknowledgement}

%
% BibTeX or Biber users please use (the style is already called in the class, ensure that the "woc.bst" style is in your local directory)
% \bibliography{name or your bibliography database}

\end{document}